# Compression properties of polymeric syntactic foam composites under cyclic loading

Z. Yousaf[1, 2*], M. J. A. Smith[1], P. Potluri[2], W.J. Parnell[1]

[1]School of Mathematics, University of Manchester, Oxford Rd, Manchester M13 9PL, UK
[2]Robotics and Textile Composite Group, Northwest Composite Centre, School of Materials, University of Manchester, Manchester M13 9PL, UK

* zeshan.yousaf@manchester.ac.uk

**Abstract**

In the present work, polymer-based syntactic foams were studied under cyclic compression in order to investigate their compressibility, recoverability, energy dissipation and damage tolerance. These syntactic foams were manufactured by adding hollow polymer microspheres of various sizes and wall thicknesses into a polyurethane matrix. The associated loading and unloading curves during cyclic testing were recorded, revealing the viscoelastic nature of the materials. SEM images of the samples were obtained in order to study potential damage mechanisms during compression. It was observed that these syntactic foams exhibit high elastic recovery and energy dissipation over a wide range of compressional strains and the addition of polymer microspheres mitigates the damage under compressional loading.

1.  **Introduction**

Syntactic foams are composite materials with widespread applications in the aerospace, automotive, sports and subsea industries due to their light weight nature, excellent acoustic properties and their buoyant behaviour [1-3]. Due to their closed-cell structure, the water absorbency of syntactic foams is significantly lower than that of the matrix material and open cell foams [4, 5]. Syntactic foams are manufactured by adding hollow thin-walled particles, known as microspheres or microballoons, into the matrix material [6]. The mechanical properties can be tailored by selecting an appropriate combination of matrix materials and hollow microspheres [7]. In the published literature, hollow microspheres of different materials namely glass, ceramic and polymers etc. have been used [8-17] with a variety of polymer matrix materials [10, 18-22]. Microspheres of different mean-wall thicknesses and diameters in varying volume fractions have also been used in matrix materials to tune the properties of the syntactic foams [3, 4, 18, 23-26]. Depending on the practical application, syntactic foams undergo varied modes of loading (e.g., compressive, tensile and shear loadings). Extensive research work has been published on syntactic foams under these loadings [3, 4, 10, 12, 24, 27-33]. Syntactic foams comprising glass microspheres are widely used for lightweight applications and have received considerable attention in subsea applications [34] . It has been reported that the stress-strain curve of glass-based syntactic foams under compression can be classified into three distinct regions, namely: the linear elastic, plateau and densification regions [4, 12, 35]. The linear region corresponds to the elastic deformation of the syntactic foam. The plateau region, in which the change in stress is insignificant, corresponds to the catastrophic failure of the glass shell due to crushing of the microspheres. The densification region is associated with the cavity filling up with debris and the matrix experiencing post-crushing. In this region, the stress increases significantly with very little change in strain [12, 35]. The main disadvantage of syntactic foams comprising glass-based microspheres is that they are stiff, brittle and are prone to damage when they are exposed to large strains and are therefore *not recoverable* [18, 19]. That is, in applications where recoverability and large strain damage tolerance are important, glass based microspheres are inappropriate. In contrast, hollow polymer microspheres offer reduced density, low price and are both softer and less brittle [19, 36]. Axial compression of individual microspheres has revealed the phenomenon of initial cell wall flattening followed by cell wall buckling [37]. This behaviour is completely different to the deformation of glass microspheres where fracture of the microspheres has been observed [38]. Compared to glass based microspheres, relatively limited research has been published on the mechanical performance of polymeric syntactic foams comprising Expancel microspheres embedded in a polymer matrix [13, 19, 39-44]. Furthermore, relevant literature on the compressibility, recoverability, damage tolerance and energy dissipation of these types of syntactic foams is scant. In this work therefore, we develop syntactic foams using polyurethane as the matrix material and polymer hollow microspheres (Expancel) as the filler. Microspheres with varying mean diameters and wall thicknesses were introduced into the polyurethane matrix, giving rise to foams with varying volume fractions of microspheres, in order to investigate the effect of mean diameter, wall thickness and volume fractions on the uniaxial compression properties of syntactic foams. Cyclic compression testing was conducted on an Instron testing machine to record the stress-strain curves of the syntactic foams. Loading and unloading curves were recorded on both virgin samples and samples that had previously undergone deformation, in order to study the compressive behaviour and time-dependent recovery of the materials. The energy dissipation and Young's modulus were also calculated for these samples. Additionally, SEM images of the tested samples were captured in order to study the damage mechanisms during compression.

## 2. Material and mechanical testing

### 2.1 Material

Polyurethane syntactic foams were fabricated by introducing hollow polymer microspheres (Expancel from AkzoNobel Sweden) into a polyurethane elastomeric matrix. The matrix material was selected and formulated to provide a high degree of confidence that microspheres could be incorporated and dispersed uniformly in the matrix, whilst having sufficiently low viscosity to allow removal of any air entrapped in the samples during mixing.

The polyurethane was formulated from a blend of Polytetramethylene Ether Glycol (PTMEG) (Terathane 1000 supplied by INVISTA Textile (UK) Ltd), Trimethylolpropane (TMP) (Tokyo Chemical Industry) and cured with Methylene diphenyl diisocyanate (MDI) (Isonate M143 - Dow Chemicals). Fumed silica (Aerosil 200 – Evonik Inc.) was used as a thixotropic additive. The matrix formulation was kept identical for all samples over the whole range of volume fractions. Hollow copolymer microspheres were incorporated into the matrix at a range of nominal volume fractions from 0% to 40%. Two types of microspheres (551 and 920 grades) with different mean wall thickness and diameter were introduced into matrix materials to study their influence on syntactic material properties under compression. The required weight of microspheres for each sample was calculated using the measured density of the 0% sample. The microspheres comprise a shell of acrylic copolymer, enclosing a fluorocarbon. Details of the microspheres 551 and 920 are presented in Table 1, and scanning-electron microscope (SEM) images of the microspheres and syntactic foams can be found in Figures 1 & 2 respectively. All ingredients were thoroughly dried and degassed before mixing; the mixed materials were cast as sheets in open trays and cured at $55^{o}$C, and after curing samples for testing were cut from the cast sheets as per standard size (to be described in the section to follow). Microspheres were added in 2%, 10% and 40% by volume in the matrix material and obtained densities of the resulting syntactic foams are presented in Fig 3, where the usual standard linear decrease in the densities of the syntactic foams with increasing microsphere volume fraction was observed.

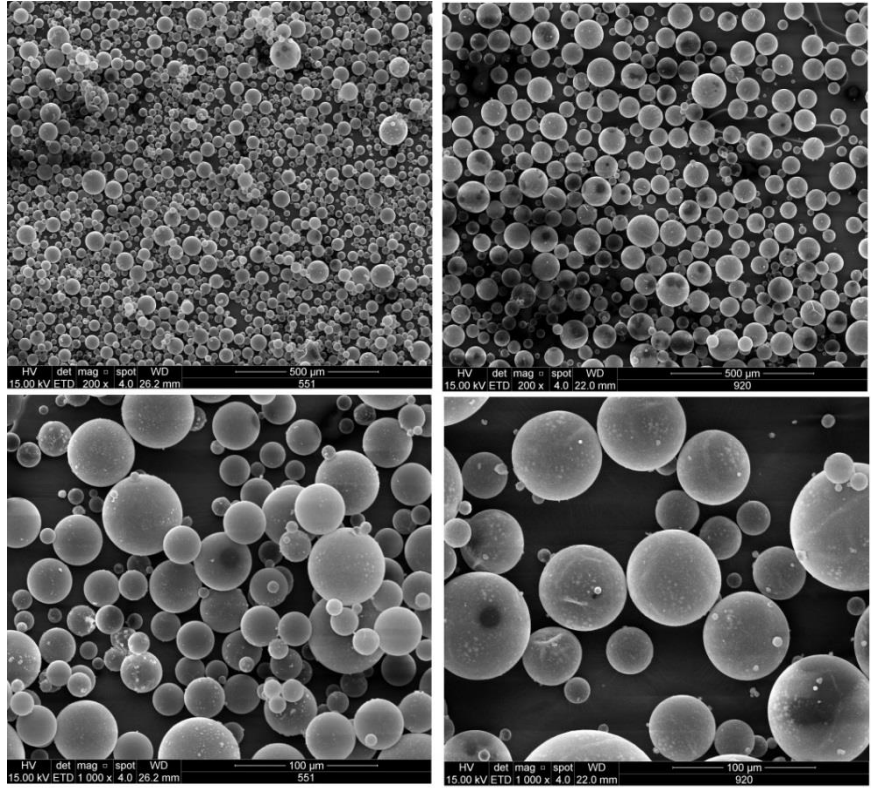

**Fig. 1.** SEM images of microsphere grades 551 (left) and 920 grade (right) at different magnifications

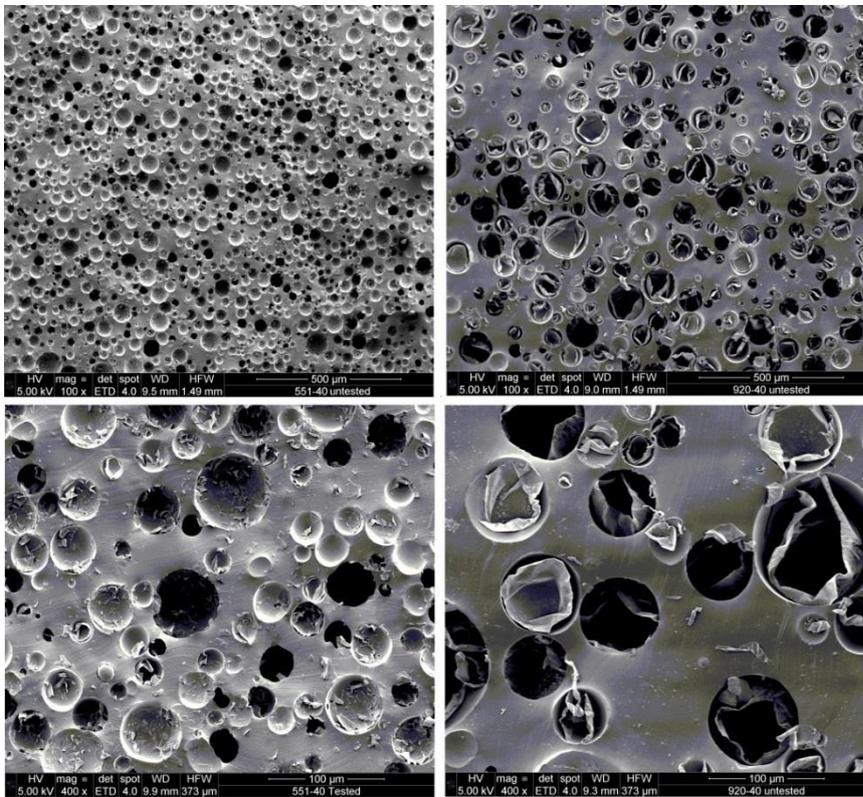

**Fig. 2.** SEM images of syntactic foams 551-40% (left) and 920-40% (right) at different magnifications

| Microsphere type | Microsphere diameter (micron) | Shell thickness (micron) | Density (g/cm$^3$) | Wall thickness-to-diameter ratio |
|---|---|---|---|---|
| 551 DE | 40 | 0.25 | 0.042 | 0.00625 |
| 920 DE | 70 | 0.35 | 0.030 | 0.00500 |

**Table 1** Details of hollow microspheres

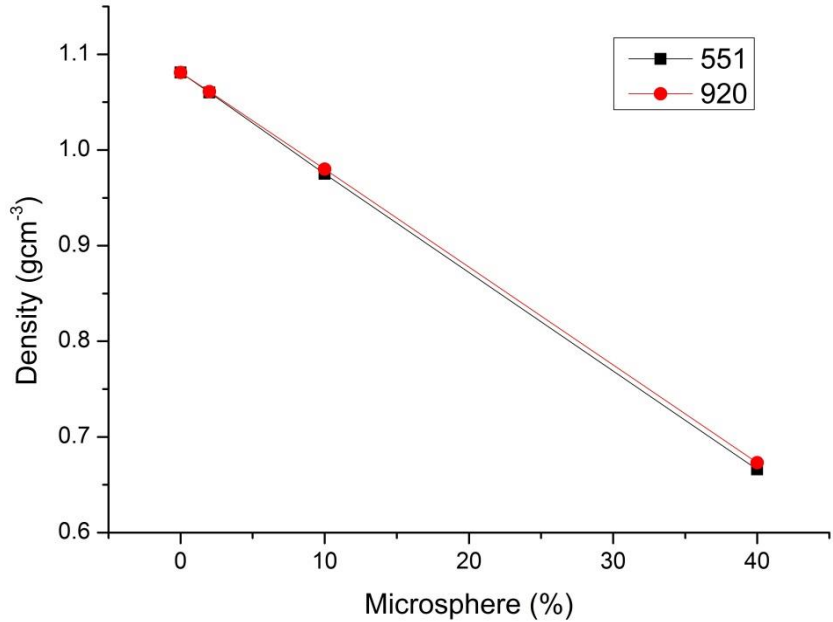

**Fig. 3.** Density variation of syntactic foams with increasing microsphere volume fraction (%)

*2.2 Mechanical testing*

Uniaxial compression tests on the syntactic foam samples were conducted on an Instron universal testing machine (Fig. 4) equipped with a 100 kN load-cell. The BS ISO 7743-2011 standard for compression testing of vulcanized rubber was followed. The samples were cut to a cylindrical shape with a diameter of 29mm and a height of 12.5mm. The samples were subjected to cyclic compression loading between flat platens with a cross-head speed of 10 mm/min. Loading and unloading curves were recorded at the same strain level (0.0132s$^{-1}$). Prior to compression, both the top and bottom platens were sprayed with a lubricant (WD-40) to minimize friction between the platens and the samples.

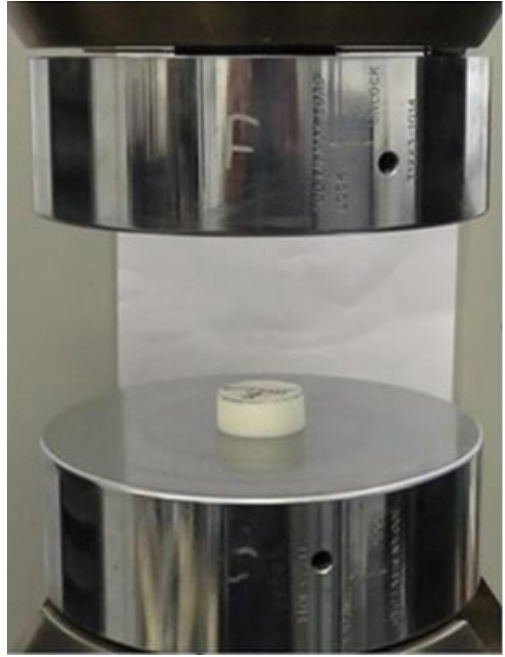

**Fig. 4.** Uniaxial compression testing set up for syntactic foam samples

## 3. Results and discussion

3.1 Macroscopic deformation

Cyclic uniaxial compression testing was conducted on all samples to 25%, 50% and 70% strains successively. Initially, cyclic loading was applied to virgin (untested) samples up to 25% strain, and after one week, cyclic testing was repeated on these same samples. Thereafter, a similar procedure was adopted for the samples up to 50% strain. However, for compression testing up to 70% strain, given that some samples were damaged (especially unfilled and syntactic foams with low microsphere concentrations) after the first five cycles (loading and unloading), cyclic testing was not repeated at 70% strain on these samples. Fig. 5 presents the pattern of the cyclic compression adopted. The thicknesses of all samples were also measured before conducting compression testing to each strain level.

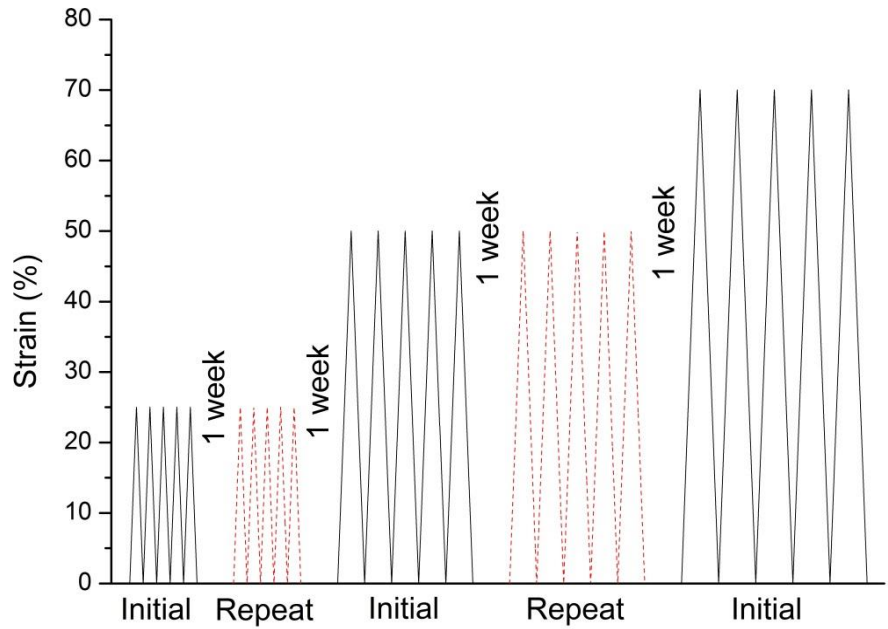

**Fig. 5.** Pattern adopted for cyclic compression

The loading and unloading curves for both unfilled polyurethane and syntactic foam samples are presented in Fig. 6. Here, the syntactic foams consisting of different microspheres (551 & 920) and volume fractions (2%, 10% & 40%) are represented by microsphere type along with the volume fraction used. The initial testing carried out on virgin samples is represented by solid lines in Fig. 6 while repeated cycles are represented by dotted lines.  At this 25% strain level, the stress-strain curves of the unfilled samples followed a non-linear behaviour typically associated with soft polymer materials [45]. When the samples were unloaded at the same strain rate as the loading cycle, an unloading curve with a different path from the loading curve was recorded, forming a hysteresis loop describing the energy dissipation during cyclic loading [46]. For the unfilled polyurethane, all five loading and unloading curves followed the same loading-unloading path as the first cycle, showing that the material returned to its original dimensions upon unloading. For both 551 and 920 grade syntactic foams with 2% and 10 % microsphere concentration, the stress-strain curves also followed a non-linear pattern similar to that of the matrix material, which reflects the dominance of the matrix material during compression at small microsphere concentrations.

For syntactic foams with 2% microsphere concentration, again all five loading and unloading curves followed the same path as the first loading-unloading curve. However, for syntactic foams with 10% concentration, after the first loading-unloading curve, the behaviour of successive loading-unloading curves was dissimilar to unfilled and syntactic foams with 2% microsphere concentration. At 10% concentration, after the first loading cycle, the subsequent four loading curves deviated from the path of the first loading curve. These four subsequent successive loadings curves followed a similar path and a smaller force was needed to induce the same strain in the last four loading curves compared to the first loading curve. This behaviour in cyclic loading, where a smaller stress is needed for reloading after first loading cycle, can be attributed to stress softening of the material [47, 48].

Both 551 and 920 grade syntactic foam samples with 40% microsphere concentration exhibited extremely different behaviour to the lower volume fraction samples, namely, the emergence of an initial linear region followed by a non-linear stress-strain response. This small initial linear region is associated with an increase in the initial stiffness of the syntactic foams with a higher concentration of microspheres. In a similar manner to the 10% concentration foams, the stress-strain curves of syntactic foams with 40% concentration also exhibited the phenomenon of stress softening in the last four loading cycles with the shrinkage of the hysteresis curve after the first loading- unloading cycle. In fact, stress softening was even more prominent in 40% concentration samples as compared to the 10% concentration samples. Unlike the unfilled and 2% concentration, it was also observed that 10% and 40% concentration syntactic foams did not return to their original configurations after the first loading cycle, exhibiting a residual strain. This residual strain can be attributed to the slow recovery of the microspheres after the first loading cycle.

After conducting cyclic testing to 25% strain on the virgin samples, the thicknesses of all of the samples were measured. Before measuring the thicknesses, the samples were left to fully relax for one week. It was observed that the dimensions of all the tested samples were unchanged, showing that the phenomenon of residual strain is fully reversible. Having examined the initial cyclic response of our samples to 25% strain, we subsequently examined the repeated cyclic curves to 25% strain after one week. The repeated cyclic testing results did not show any noticeable change compared to those for the virgin samples. The repeated curves for the unfilled and syntactic foam samples are represented by dotted lines in Fig. 6. Again, the phenomenon of stress softening and residual strain was apparent, as is indicated in the stress-strain graphs for 10% and 40% concentration. After testing the previously strained specimens, the thicknesses of all the samples were measured again and it was observed that changes in dimension were insignificant: only a 1% change in the thickness was recorded for syntactic foams with 40% concentration while the other samples did not show any measurable change in thicknesses.

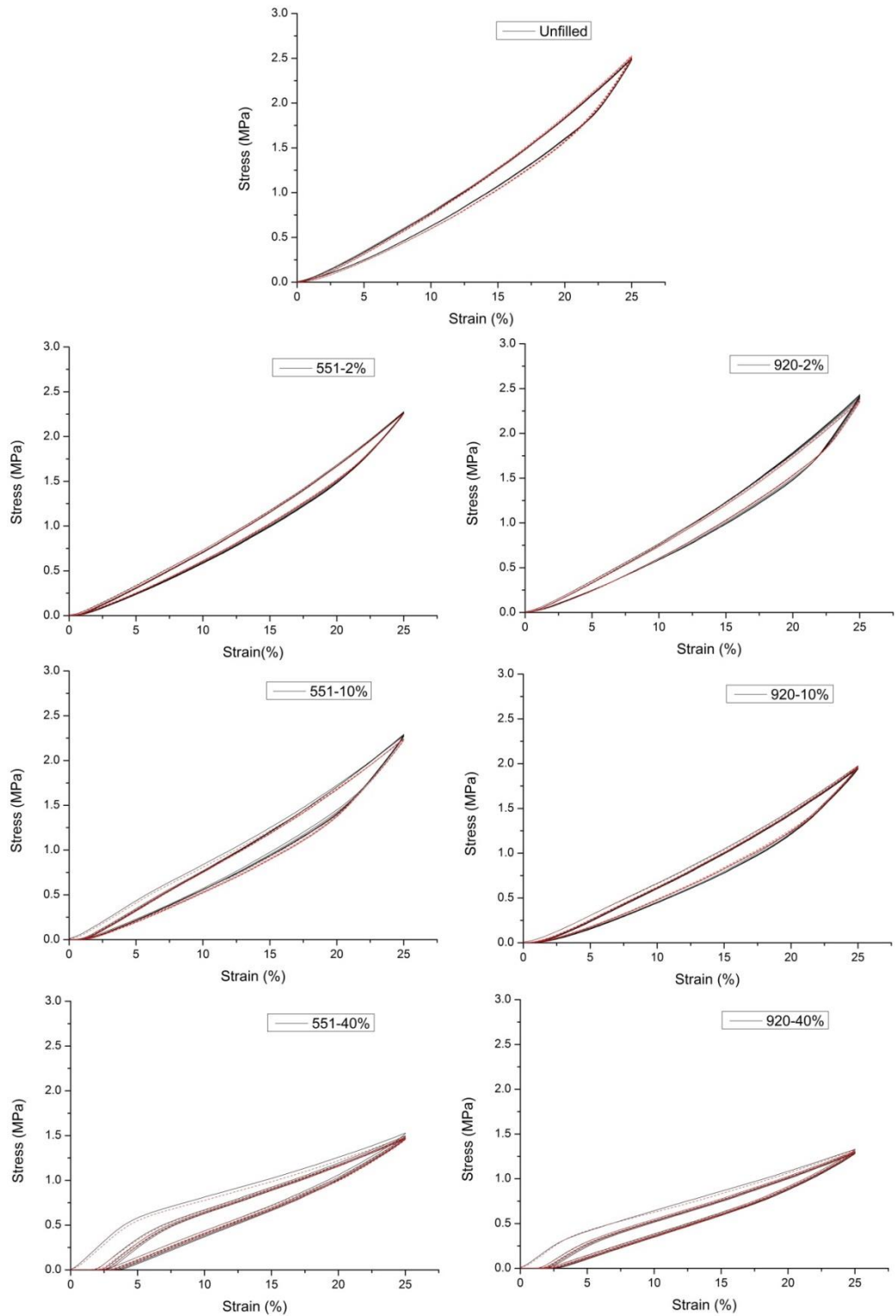

**Fig. 6.** Stress-strain curves for unfilled, 551 and 920 syntactic foams up to 25% strain. Initial testing cycles are represented by solid lines (black colour) and repeated testing cycles by dotted lines (red colour). Strong recoverability after initial testing is highlighted, given that the repeat tests follow, almost exactly, the initial loading curves.

After performing cycling testing on virgin and pre-tested samples up to 25% strain, the same specimens were tested to 50% strain. The stress-strain response of the specimens up to 50% strain levels are presented in Fig. 7 with solid lines representing initial stress-strain results and dotted lines showing repeated stress-strain results. The stress-strain curves for the unfilled and syntactic foam present similar trends to those observed in the testing of samples up to 25% strain (with hysteresis loops for all samples). Here again, we observed a non-linear stress-strain behaviour for unfilled and syntactic foams with 2% and 10% volume fractions, and a small linear region for syntactic foams with 40% concentration. The phenomenon of stress softening and residual strain was also clearly evident for syntactic foams with 10% and 40% microsphere concentrations. The repeated test results for stress-strain up to 50% (expressed by dotted lines in the same graph) followed a similar path to the initially tested samples with very little deviation in the loading-unloading curves. The thicknesses of the samples were measured after initial 50% strain testing and repeated 50% strain testing; the thicknesses of the syntactic foams with 2% and 10% concentrations was unchanged while an insignificant change (around 1-1.2%) was recorded for syntactic foams with 40% microsphere concentration. Reasons for this thickness reduction include the fact that the microspheres have not fully recovered for this high concentration (due to either extremely slow stress relaxation or plastic deformation).

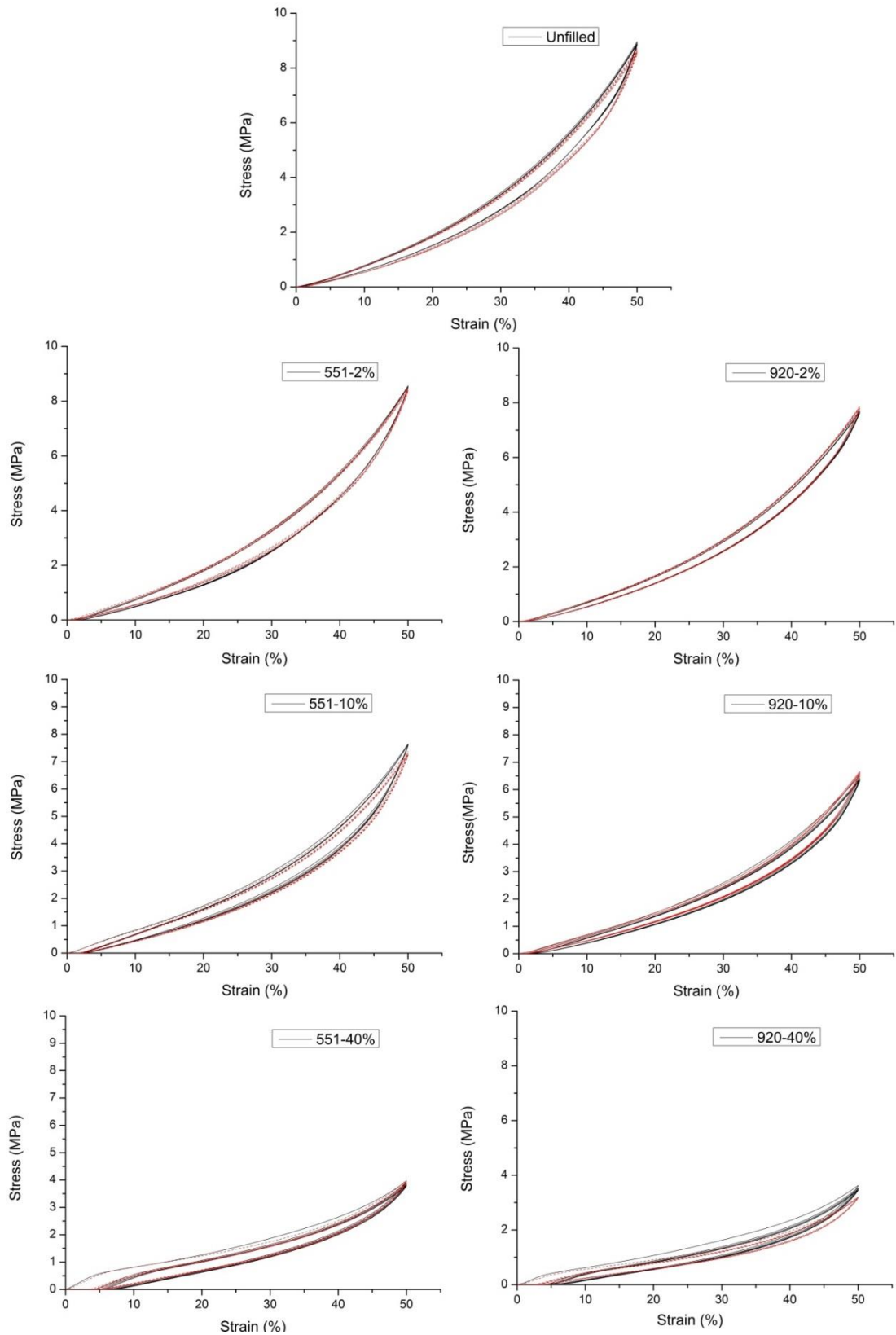

**Fig. 7.** Stress-strain curves for unfilled, 551 and 920 syntactic foams up to 50% strain. Initial testing cycles are represented by solid lines (black colour) and repeated testing cycles by dotted lines (red colour)

After repeat compression testing to 50% strain level, samples were left to fully relax for one week before proceeding to compression testing at the next strain level, as described earlier (see Fig. 5). Finally, the samples were tested to 70% strain in order to study the response of these materials at a level of high strain. As discussed earlier, cyclic stress-strain data for this strain level was not repeated as samples started to break at this strain level. The stress-strain results up to 70% strain are presented in Fig. 8 along with the tested sample images in the same figure. Here also the stress-strain curves were non-linear for unfilled and syntactic foams with 2% and 10% microspheres as observed for 25% and 50% strain levels. The stress- strain curve for 40% microspheres showed the general trend of glass-based syntactic foams in the literature with an initial linear region followed by a plateau region and finally a densification region [12, 49]. The peak stresses in the initial linear region for Expancel-based syntactic foams obtained in the present study are much smaller than that for glass-based syntactic foams observed in the literature [12], due to the lower stiffness of Expancel microspheres compared to glass microspheres. However, in contrast to the stress-strain curves of glass-based syntactic foams where a distinct yield point was observed at the end of the elastic region, no such yield point was present on the stress-strain curves of polymer based syntactic foams. We speculate that this difference is due to the fact that polymeric microspheres do not break but instead buckle, in contrast to the glass microspheres [12, 37]. Similarly in the plateau region of the stress-strain curves of polymer based microspheres, we hypothesize that the microspheres continue buckling instead of crushing; when the upper and lower shell walls touch, the densification region appears for these syntactic foams. Returning to Fig. 8, the stress-strain curves for unfilled and syntactic foams with 2% microspheres were not smooth but showed kinks and a significantly high drop in peak stress values, especially after the first loading curves. This drop in stress corresponds to sample breaking, with damage visibly apparent at the macro-level on inspecting the tested samples. The curves for syntactic foams with 10% and 40% microsphere concentration were fairly smooth and did not show any kink/sudden load drop. Only an insignificant decrease in peak stresses was observed during the last loading cycles for syntactic foams at higher microsphere concentrations. The results were also supported from tested sample images at the macro-scale (inset in Fig. 8). As can be seen from the tested sample images, the unfilled samples were badly damaged while damage decreased progressively with increasing microsphere concentration. No apparent damage was present in syntactic foams with 40% microsphere concentration. Interestingly, the opposite trend has been reported for glass-based syntactic foams where increased macro-scale damage was observed for *higher* concentrations of glass microspheres, due to the brittle nature of these microspheres [18, 25, 35].

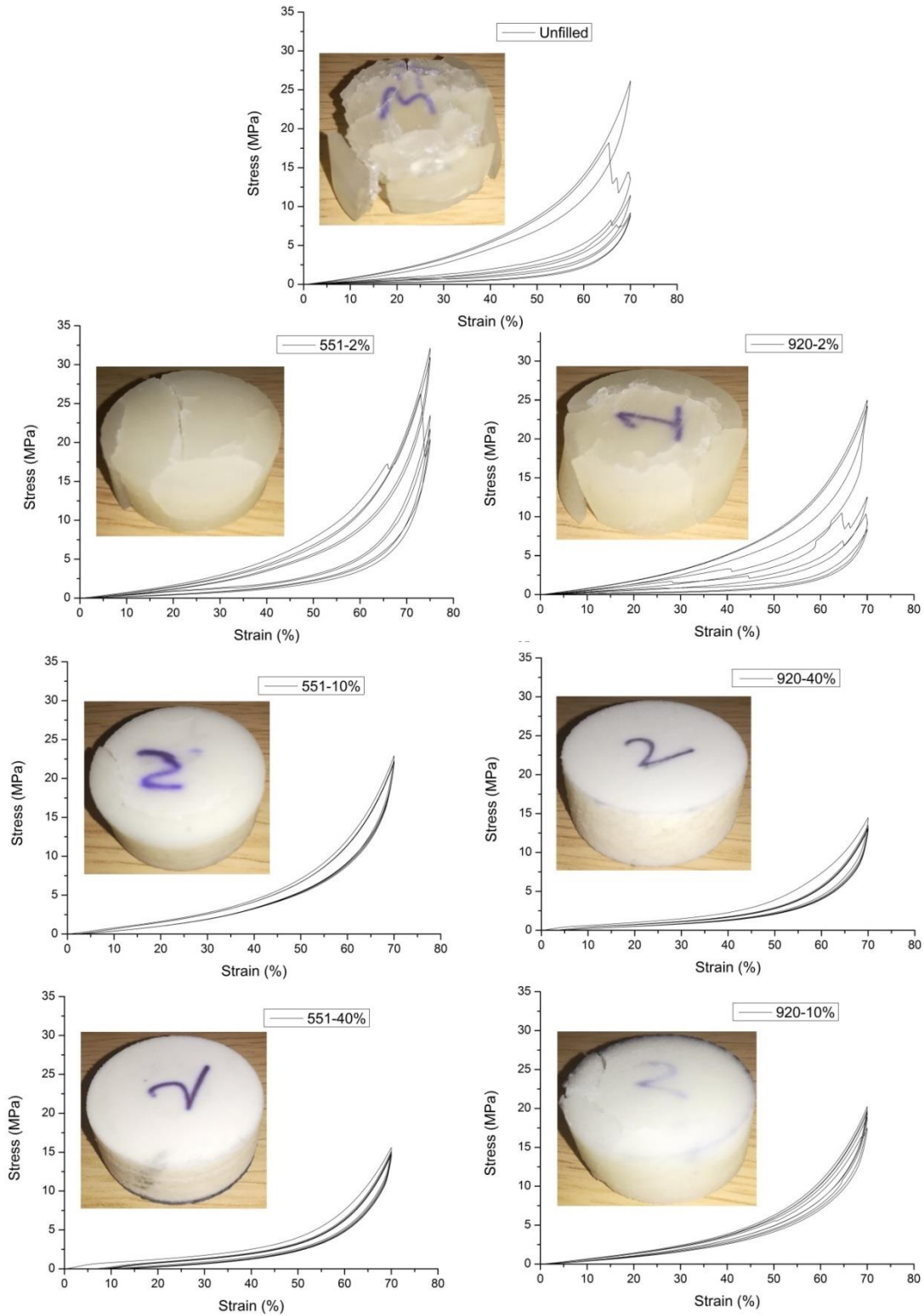

**Fig. 8.** Stress-strain curves for unfilled, 551 and 920 syntactic foams up to 70% strain. Tested samples to 70% strain are embedded inside the graphs. At this high strain level, macro-scale damage appears to decrease with an increase in microsphere volume fraction in contrast to glass-based syntactic foams.

3.2 Microscopic deformation

After compression testing the unfilled and syntactic foams up to 70% strain, SEM images of these samples were obtained in order to study the damage of the samples at the micro-scale. SEM images of tested samples are presented in Fig. 9 and Fig. 10, where it was observed that the damage was highest in unfilled samples with visible matrix cracking after loading. In syntactic foam samples, it was observed that cracks propagated through the matrix and the presence of polymer microspheres acted as a barrier for damage propagation and prevented crack front propagation. These images also explain the load drop observed at the macro-scale during the mechanical testing of samples, demonstrating almost certainly that the load drops are due to sample breakage.

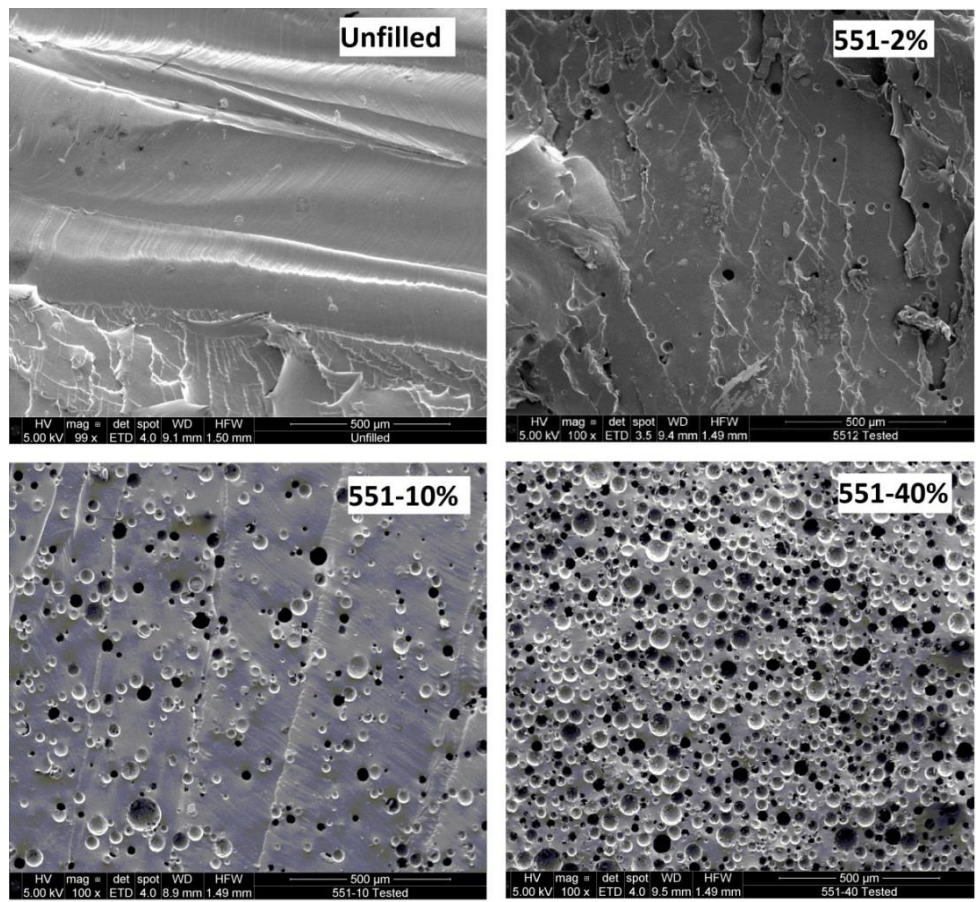

**Fig. 9.** SEM images of unfilled and 551 syntactic foams after 70% strain. Damage is mitigated in high volume fraction foams by the presence of microspheres.

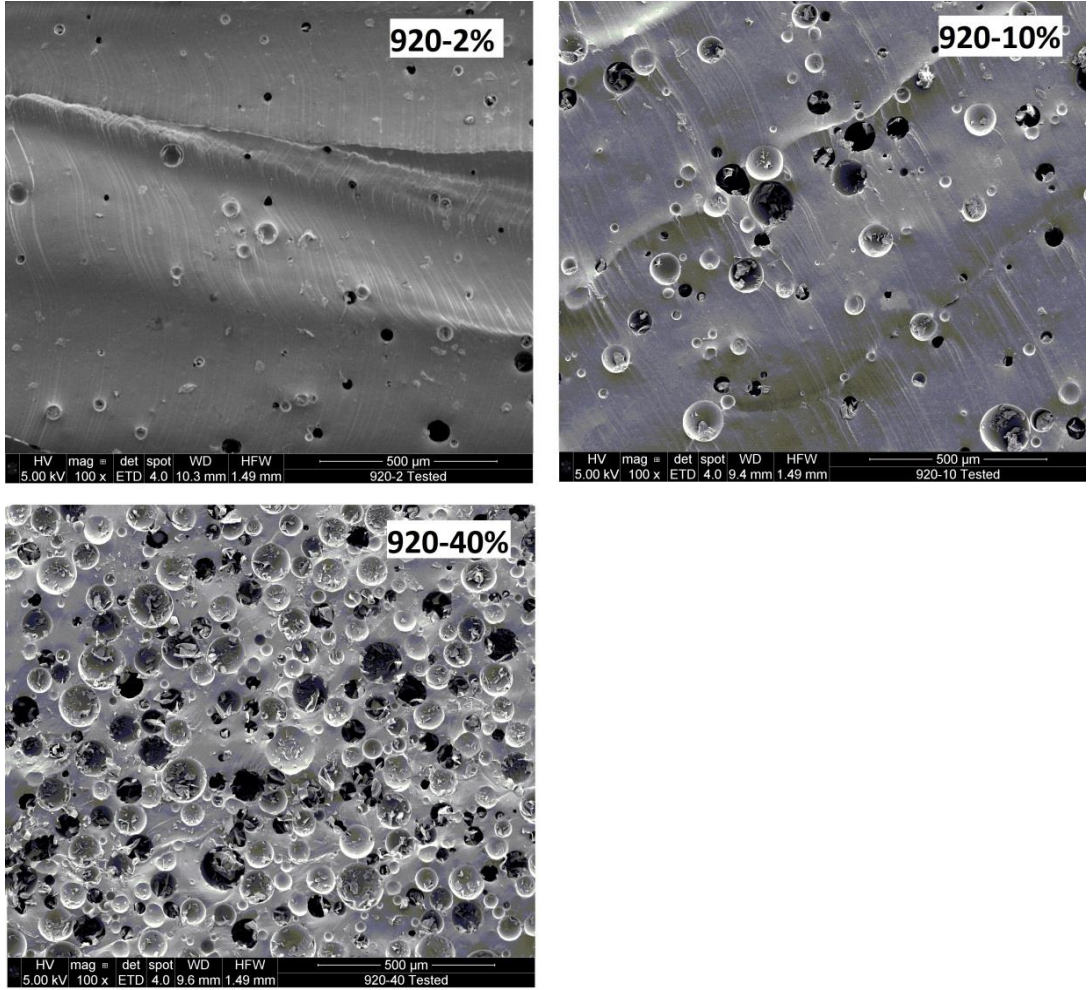

**Fig. 10.** SEM images of 920 syntactic foams after 70% strain. Damage is mitigated in high volume fraction foams by the presence of microspheres.

3.3   Energy dissipation of unfilled polyurethane and syntactic foams

During the cyclic testing of unfilled polyurethane and syntactic foams, it was observed that a hysteresis loop was formed during the loading and unloading process. After determining the stress-strain results of unfilled polyurethane syntactic foams under cyclic loading, we computed the energy dissipation associated with these samples for 25% and 50% strain levels. The dissipated energy $W_{diss}$ for all five cycles was calculated using the expression [50]

$$W_{diss} = \int_0^t \left(\sigma^L(t) - \sigma^U(t)\right)\frac{d\varepsilon}{dt}dt = \int_{\varepsilon(0)}^{\varepsilon(t)} \{\sigma^L(\varepsilon) - \sigma^U(\varepsilon)\}d\varepsilon \qquad (1)$$

where $\sigma^{L,U}$ denotes the engineering stress under loading and unloading, respectively, and $\varepsilon$ denotes the engineering strain. The integral above was evaluated straightforwardly using a trapezoidal rule, i.e.

$$W_{diss} \approx \sum_{k=1}^{N} \left( \frac{\sigma_{k-1}^L + \sigma_k^L}{2} - \frac{\sigma_{k-1}^U + \sigma_k^U}{2} \right) \Delta \varepsilon_k \qquad (2)$$

where $\Delta \varepsilon_k$ is the width of the *k*-th subinterval. The energy dissipation versus the number of cycles for the unfilled polyurethane and both grades of syntactic foams up to 25% strain is presented in Fig. 11 while energy dissipation up to 50% strain is depicted in Fig. 12. The energy dissipation for these samples was highest for the first cycle (with the exception of syntactic foams with 2% microsphere volume fraction). The energy dissipation was also higher in syntactic foams with higher microsphere volume fractions (10% and 40% microsphere volume fractions), with 551 grade syntactic foams at 40% microsphere volume fraction dissipating the greatest energy. For the first cycle, 551-40% syntactic foams dissipated 76% and 37% more energy than unfilled and 920-40% syntactic foams, respectively at 25% strain level, while at 50% strain level, the energy dissipation for 551-40% syntactic foams was 82% and 12% higher than unfilled and 920-40% syntactic foams. After the first cycle, the energy dissipation for syntactic foams 551 and 920 with 10% and 40% microsphere volume fractions decreased sharply while the change in energy dissipation for the successive cycles was very small. The change in energy dissipation with increasing number of cycles for unfilled polyurethane and syntactic foams with 2% microsphere was not very sharp. This was also evident from the shrinkage of the hysteresis loop for syntactic foams containing higher volume fractions of microspheres of both grades (551 and 920), while the change in the hysteresis loops for unfilled and 2% microsphere volume fraction was not noticeable with successive cycles. Energy dissipation is the result of various factors, e.g. the viscoelastic behaviour of the material, debonding between the microspheres and the matrix, and general damage of the structure [51] . Due to the strong viscoelastic nature of syntactic foam materials, we believe the energy dissipation is mainly due to viscoelastic effects as higher energy dissipation and minimal damage was observed for syntactic foams with higher microsphere volume fractions.

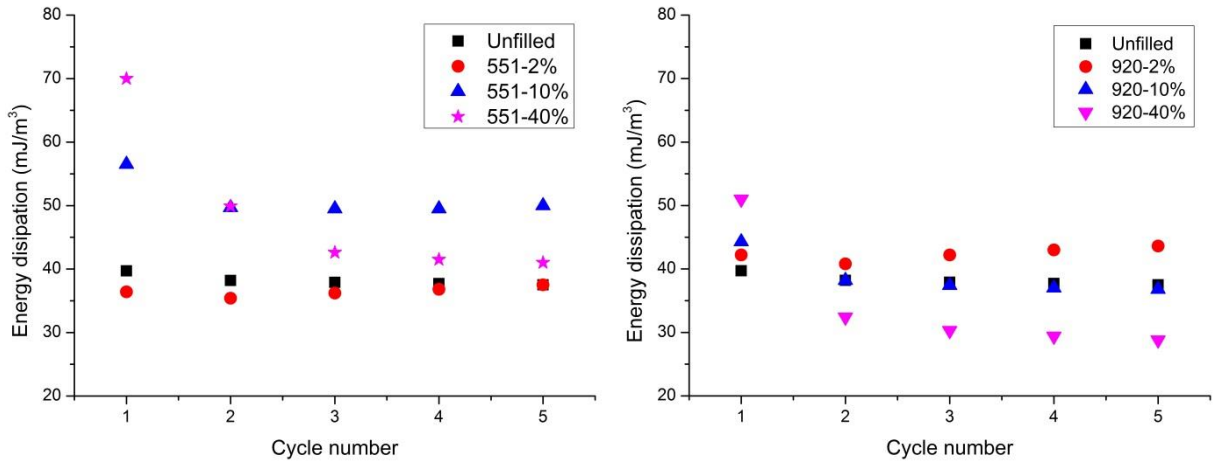

**Fig. 11.** Energy dissipated per unit volume of unfilled polyurethane and syntactic foams versus number of cycles up to 25% strain

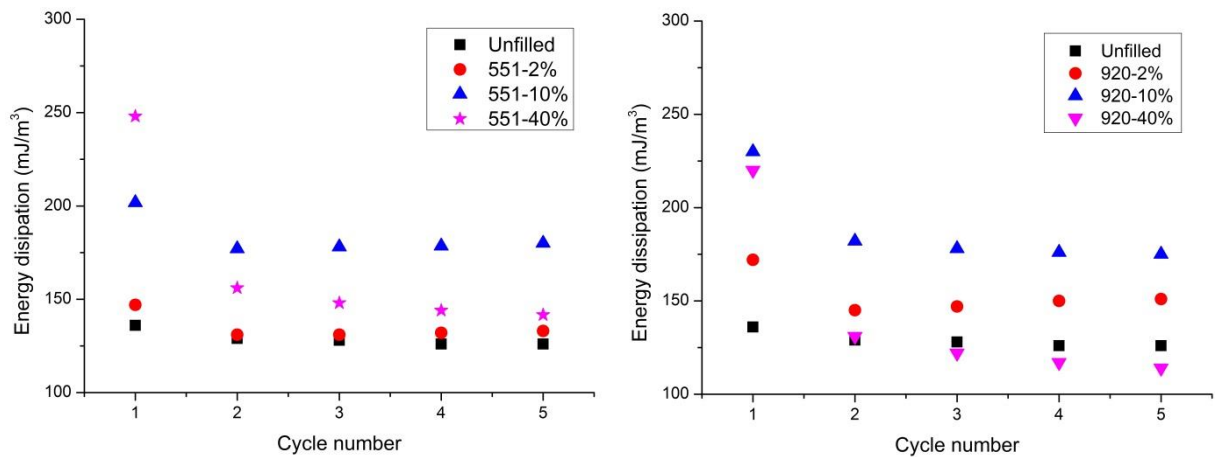

**Fig. 12.** Energy dissipated per unit volume of unfilled polyurethane and syntactic foams versus number of cycles up to 50% strain

3.4   Modulus of unfilled polyurethane and syntactic foams

During compression testing, it was observed that the initial stiffness of syntactic foams was higher than the unfilled polyurethane at small strain levels. That is, the elastic modulus increased with an increase in microsphere volume fraction. It was noted that the modulus of 551 syntactic foams was higher than for 920 syntactic foams (Fig. 13). At 40% microsphere volume fraction, the elastic modulus of syntactic foams containing 551 microspheres was 32% higher than for 920 syntactic foams at the same volume fraction. We hypothesize that this increase is due to the higher wall thickness-to-diameter ratio in 551 syntactic foams. A similar effect was reported in glass-based syntactic foams [4, 18, 25] where the authors observed an increase in the initial stiffness and peak stresses for syntactic foam containing thicker wall microspheres.

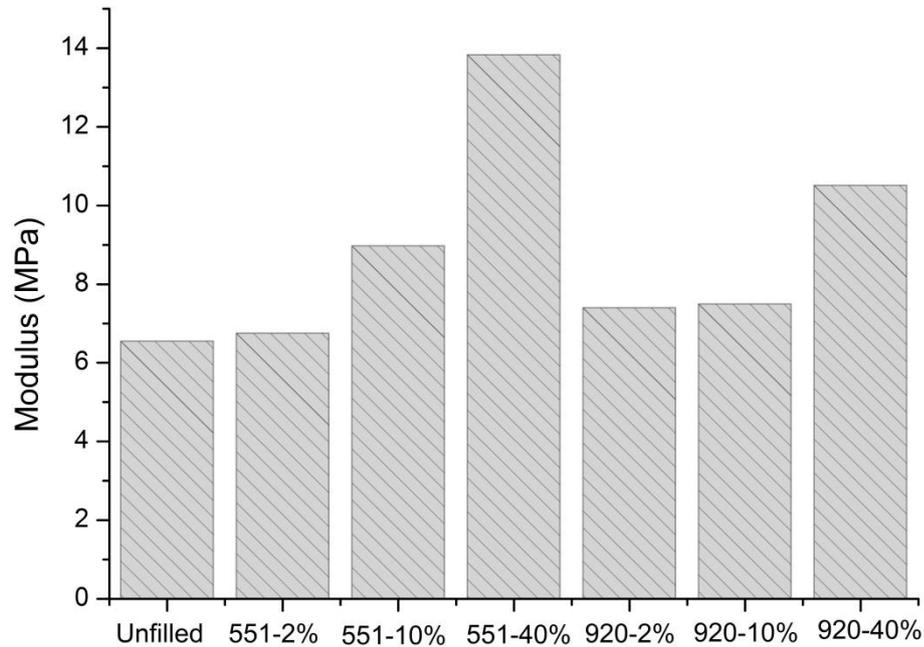

**Fig. 13.** Elastic modulus of unfilled polyurethane and syntactic foams

**4. Conclusions**

We have conducted a comprehensive study on unfilled polyurethane and polyurethane based syntactic foams comprising polymer microspheres under cyclic uniaxial compression. Five cycles (loading and unloading) were performed on each sample up to strains of 25%, 50% and 70%. Cyclic compression testing on these materials was conducted twice, with secondary cyclic testing conducted after one week, in order to study their time-dependent recovery and compression properties. The stress-strain results revealed the non-linear behaviour of unfilled polyurethane and syntactic foams with smaller microsphere volume fractions. Hysteresis curves were recorded for unfilled polyurethane and syntactic foam samples revealing the viscoelastic nature of these materials. In addition to hysteresis, the phenomenon of stress softening and residual strain was also observed for the successive loading-unloading curves in syntactic foams with higher volume fractions. The behaviour of syntactic foams with higher microsphere volume fractions was found to be qualitatively similar to glass-based syntactic foams, as the stress-strain curve could be divided into three regions, namely the linear region, plateau region and finally a densification region. The stress-strain curves showed that the initial stiffness and consequently the elastic modulus of these syntactic foams increased with an increase in the microsphere concentration. These polymeric syntactic foams showed excellent elastic recovery after the removal of load and a significant reduction in damage with increasing the volume fraction of polymer microspheres, compared to other syntactic foams (e.g. glass and ceramic based). Damage was mitigated with higher volume fractions of polymer microspheres in these syntactic foams contrary to other syntactic foams at similar volume fractions. Results were supported by SEM images of the fractured surface where most of the damage was observed in the form of matrix cracks. It was also observed that the wall thickness-to-diameter ratio influenced the compressional properties of the syntactic foams with higher modulus and peak loads obtained with syntactic foams possessing microballons with higher wall thickness to diameter ratios. Additionally, the syntactic foams with higher wall thickness-to-diameter ratio dissipated greater energy than unfilled and syntactic foams with lower wall thickness-to-diameter ratios. Based on these results it is possible to further tailor the properties (density, initial stiffness, peak load, energy dissipation and fracture strain) of these syntactic foam polymer composites by changing the microsphere volume fractions and wall thickness-to-diameter ratios for applications requiring low density, recoverability, high damage tolerance and energy dissipation.

**Acknowledgements**

We acknowledge Alison Daniel ( Thales UK) for sample manufacturing. The authors would also like to acknowledge EPSRC (grant EP/L018039/1) and Thales (Thales Research Structures, Materials and Acoustics Research Technologies (SMART) Hub) for funding this project.